\documentclass[prd,letter,eqsecnum,nofootinbib,preprint,floatfix]{revtex4}
\usepackage{colordvi}
\usepackage{graphicx}
\usepackage{bm}
\usepackage{multirow}
\usepackage{enumerate}
\reversemarginpar
\def\<{\langle}
\def\>{\rangle}
\newcommand{\text}{\rm}

\def\tr{{\text tr}\,}
\def\Eq#1{Eq.~(\ref{#1})}

\def\Sec#1{Section~\ref{#1}}
\def\Ref#1{Ref.~\cite{#1}}

\begin{document}

\begin{titlepage}

\vspace*{0.7in}
 
\begin{center}
{\large\bf Partial breakdown of center symmetry in  large-$N$ QCD\\ with adjoint Wilson fermions.
}
\vspace*{1.0in}

\vspace{-1cm}
{Barak Bringoltz\\
\vspace*{.2in}
Department of Physics, University of Washington, Seattle,
WA 98195-1560, USA
}
\\
and\\
Department of Particle Physics, Weizmann Institute of Science, Rehovot 76100, Israel
\vspace*{0.25in}

\end{center}

We study the one-loop potential of large-$N$ QCD with adjoint Dirac fermions. Space-time is a discretization of $R^3\times S^1$ where the compact direction consists of a single lattice site. We use Wilson fermions with different values of the quark mass $m$ and set the lattice spacings in the compact and non-compact directions to be $a_t$ and $a_s$ respectively. Extending the results of JHEP 0906:091,2009, we prove that if  the ratio $\xi = a_s/a_t$ obeys $0<\xi<2$, then the minimum of the one-loop lattice potential for one or more Dirac flavors is $Z_N$ symmetric at the chiral point. For $\xi=0$ our formulas reduce to those obtained in a continuum regularization of the $R^3$, and our proof holds in that case as well.
As we increase $m$ from zero, we find a cascade of transitions where $Z_N$ breaks to $Z_{K}$. For very small masses, $K\sim 1/(a_tm) \gg 1$, while for large masses $K\sim O(1)$. Despite certain UV sensitivities of the  lattice one-loop potential, this phase structure is similar to the one obtained in the continuum works of Kovtun-\"Unsal-Yaffe, Myers-Ogilvie, and Hollowood-Myers. We explain the physical origin of the cascade of transitions and its relation to the embedding of space-time into color space.

\vspace*{0.15in}

\end{titlepage}

\setcounter{page}{1}
\newpage
\pagestyle{plain}

\section{Introduction}
\label{intro}
There is much recent interest in the phenomenon of space-time reduction of $SU(N)$ gauge theories \cite{EK}. Reasons for this include the hope to understand nonperturbative dynamics of QCD with analytic small volume techniques \cite{mithat}, and the possibility to reduce the numerical cost in calculating properties of large-$N$ gauge theories on the lattice. For related works on non-lattice simulations see relevant references in \cite{Hanada}.

The nonperturbative definition of spacetime reduction requires a nonperturbative definition of the gauge theory whose volume one wishes to reduce. This requirement naturally leads one to a lattice definition. Indeed, all the known proofs of spacetime reduction that go beyond perturbation theory involve a lattice regularization. These proofs (see for example Refs.~\cite{EK,Neuberger02,KUY1}) state the following.

\bigskip

Given an $SU(N)$ lattice gauge theory defined by a set of dimensionless couplings (the lattice `t Hooft  coupling $g^2N$, a set of bare masses in lattice units $am_{1},am_2,\dots$, etc.), and residing on a geometry defined by a set of integers $L_{1,2,\dots}$ counting the number of sites along the different compact directions, then, under certain conditions, physical observables (e.g.~Wilson loops, meson propagators, quark condensates, etc.) are independent of $L_{1,2,\dots}$ when $N=\infty$. These conditions are
\begin{enumerate}
\item The ground state is translation invariant.
\item The ground state is invariant under center transformations that rotate Polyakov loops that wrap the compact directions.
\item The ground state obeys cluster decomposition (large-$N$ factorization).
\end{enumerate}

The validity condition that is hardest to fulfill seems to be no.~(2). In the theory of interest to the current paper -- four dimensional $SU(N)$ gauge theory defined on $R^3\times S^1$ with  adjoint fermions -- this condition involves the center symmetry $Z_N$ whose order parameter is the Polyakov loop $\Omega$ that wraps the $S^1$.
The ultimate check of whether the ground state of this theory respects the center symmetry can be performed only with nonperturbative lattice simulations. Nevertheless, one can approach this question analytically in certain regimes. 

In this paper we focus on the regime of weak-coupling and use perturbation theory to calculate the effective potential that governs the expectation value of  $\Omega$. This analysis is justified when the inverse length of the $S^1$ is large compared to $\Lambda_{QCD}$, and, for the massless case,  the  calculation was performed in the continuum theory by Kovtun, \"Unsal and Yaffe in \Ref{KUY}. For nonzero masses, the continuum calculation was done by Myers and Ogilvie, and Hollowood and Myers in Refs.~\cite{MO,MH}.\footnote{While \Ref{MO} preceded \Ref{MH}, the latter used techniques that allow one to avoid a numerical minimization of the one-loop potential. Also, we note in passing that while some of the results of \Ref{MH} involve the space $S^3\times S^1$, then it is the result in the large $S^3$ limit, where the sphere becomes an $R^3$, that is relevant to our calculation. For earlier results on $S^3\times S^1$ see \Ref{Mithat2}.}

The picture emerging from these weak coupling calculations is interesting\footnote{Here we restrict to more than a single Majorana fermion. The SUSY case is more complicated and we refer to \Ref{KUY} for further discussion.}: At zero mass the ground state is $Z_N$ invariant. As the mass increases from zero there appears a cascade of transitions where the center breaks to smaller and smaller subgroups. When the mass is infinite the symmetry breaks completely.\footnote{Clearly this assumes that $N$ is non-prime (otherwise $Z_N$ would not accommodate subgroups). If $N$ is prime, then a would-be instability towards a ground state that has a $Z_K$ symmetry with $K<N$,  will actually lead to a ground state that completely breaks $Z_N$, but that `looks' approximately $Z_K$ symmetric. For example, suppose $N$ is odd and choose the mass to be such that for an $SU(N')$ gauge group with $N'=N+1$ the center $Z_{N'}$ would break to $Z_2$. This will make the eigenvalue density of the Polyakov loop double peaked. The intact $Z_2$ symmetry will ensure that the two peaks have identical heights. For the same mass the eigenvalue density in the $SU(N)$ case will also have two peaks, but of different heights. At large enough values of $N$, however, the difference between the $SU(N)$ and $SU(N+1)$ theories will be washed away and the peaks will differ by a $1/N$ amount. Thus, even the $N=$ odd case can experience an approximate $Z_N\to Z_2$ symmetry breakdown.\label{prime}}

The aim of this paper is two-fold. First, we wish to gain physical insight on the origin of this cascade of transitions. Second, we wish to generalize the calculations of Refs.~\cite{KUY,MO,MH} to the lattice regularization defined with a single site in the compact direction.
Our interest in this setup derives from the UV sensitivities of the one-loop lattice potential that we analyzed in \Ref{one_loop}. These can make the  $Z_N$ realization of the lattice theory different  than that of the continuum theory, and even dependent on the lattice action. Therefore, we choose an action which is the one simulated in \Ref{BS} -- pure Yang-Mills fields regularized with the simple plaquette Wilson action and a fermion regularization of the Wilson type. We also allow for an asymmetry between the lattice spacing in the non-compact directions, $a_s$, and the lattice spacing $a_t$ in the compact direction. The ratio of the two spacings is denoted by 
\begin{equation}
\xi=a_s/a_t.
\end{equation}

The current paper can be viewed as a continuation of \Ref{one_loop}, where we analyzed the same setup studied here, but in a more restricted way and with somewhat different methods. In particular, in \Ref{one_loop} we compared the energy of different vacua by explicitly studying  particular realizations of the eigenvalues of  $\Omega$. In contrast, in the current paper we directly investigate the eigenvalue density of $\Omega$ and the way it may realize different symmetry breakdowns. This method is advantageous to the one used in \Ref{one_loop} because it allows one to study many type of symmetry realizations. Indeed, with the relatively inefficient calculation method of \Ref{one_loop}, we only examined the energy of the ground state where $Z_N$ is intact, of the one that completely breaks $Z_N$, and of the one that partially breaks $Z_N$ to $Z_2$. In the current paper we study many more possible symmetry realizations of the form $Z_N\to Z_K$ with $K< N$, and for simplicity assume that $N/K=$ integer (but see footnote \ref{prime}). 

Our current calculation method also allows us to make certain analytic statements that do not rely on numerical evaluations of the one-loop free energy, and leads to a simple picture  of the emergent phase structure that is connected to the embedding of spacetime in color space.

The following is the outline of this paper. In \Sec{oneloop} we show how to recast the one-loop potential $V(\Omega)$ calculated in \Ref{one_loop} in a form useful to the purposes of this paper. Specifically, we write 
\begin{equation}
V(\Omega) = 2\sum_{r=1}^\infty\, V_r\, \left|\,{\rm tr}\, \Omega^r\right|^2 + {\rm constant}.
\end{equation}
Since the density $\rho$ of the eigenvalues of $\Omega$ is defined by
\begin{equation}
\rho(\omega) \equiv \left\<\frac1N\sum_{a=1}^N \, \delta \left(\theta^a - \omega\right)\right\>,
\end{equation}
then its $r^{\rm th}$ moment, $\rho_r$, defined by
\begin{equation}
\rho_r = {\rm Re}\,\int \frac{d\omega}{2\pi}\, e^{i\omega r}\, \rho(\omega),
\end{equation}
obeys
\begin{equation}
\rho_r = {\rm Re}\,\left\<\frac1N\,{\rm tr}\, \Omega^r\right\>.
\end{equation}
This means that the sign of $V_r$ determines whether the eigenvalue density develops an instability in its $r^{\rm th}$ moment and breaks the symmetry from $Z_N$ to $Z_r$.

In \Sec{minimum} we minimize $V(\Omega)$ by investigating the signs of $V_r$ for different values of $m,a_s,a_t$, and $N_f$. The mathematical  form we obtain for $V_r$ makes clear the spacetime embedding into color space, and leads one to anticipate and understand the transition cascade obtained in Refs.~\cite{MO,MH}. We discuss this point in \Sec{meaning}.
Then, we turn to minimize the one-loop potential. We start with the $m=0$ case, where we determine the sign of $V_r$ analytically: for finite values of $a_{s,t}$ see \Sec{m0_lat}, and for the $a_s\to 0$ limit see \Sec{m0_cont}.  
 We proceed in \Sec{m_nonzero} to calculate $V_r$ numerically for general values of the parameters (since $V_r$ takes the form of an integral over complicated functions defined within the lattice spatial Brillouin zone, we cannot evaluate it analytically for generic parameters). We summarize  our results in \Sec{summary}.

\bigskip

We note in passing that in a related study, \Ref{BBCS} calculated the one-loop potential of the single site theory that we study here, and regulated the UV of the $R^3$ in the continuum. As \Ref{one_loop} showed, if certain relevant operators are added to the one-loop potential of \Ref{BBCS}, then such a continuum analysis is expected to yield results which are consistent with those we report here (at least up to finite lattice corrections) and with those reported in \Ref{KUY}. Nevertheless, the results of \Ref{BBCS} indicated that the zero mass system has a ground state that breaks the $Z_N$ symmetry; a result that contrasts the results of the works mentioned above. Our current analysis shows that this $Z_N$ breaking is a result of an erroneous sign in the final expression for $V(\Omega)$ that \Ref{BBCS} minimized. For clarity, we re-derive in Appendix~\ref{appC} the form of $V(\Omega)$ with a continuum regulator for the non-compact UV and show that 
it yields results which are consistent with those appearing in \Ref{KUY} and in the current paper.

\section{The one loop potential}
\label{oneloop}
In this section we calculate the one-loop potential and recast it in a useful form for our purposes. This is done in \Sec{oneloop_calc}. In \Sec{meaning} we discuss some of its general features.

\subsection{The calculation}
\label{oneloop_calc}
In \Ref{one_loop} we calculated the one-loop potential $V$ as a function of the Polyakov loop $\Omega$ that wraps the $S^1$. In terms of the eigenvalues $e^{i\theta^a}$ of $\Omega$ the result was
\begin{eqnarray}
V(\Omega) &=&  \sum_{a\neq b} \, \int_{-\pi}^\pi \left(\frac{dp}{2\pi}\right)^3\, \left\{\log \left[ \frac1{\xi^2}\, \hat p^2 + 4\sin^2\left(\frac{\theta^a-\theta^b}{2}\right)\right]\right.\nonumber \\
&&\left. - 2N_f \log\left[   \frac1{\xi^2}\, {\hat{\hat{p}}}^2 + \sin^2\left(\theta^a-\theta^b\right) + m^2_W(p,\theta^a-\theta^b)\right]\right\}.\label{1loop}
\end{eqnarray}
To obtain this result we neglected the contribution of a set of fluctuations in the $N$ diagonal components of the gauge fields, which is justified at large-$N$.

The quantities $\hat p$ and $\hat{\hat{p}}$ are the lattice momenta 
\begin{equation}
\hat p^2 = \sum_{i=1}^3 4\sin^2 \left(p_i/2\right),\qquad \hat{\hat{p}}^2 = \sum_{i=1}^3 \sin^2 p_i,
\end{equation}
and $m_W$ accounts for the  contribution of the bare mass  and the Wilson term to the propagator
\begin{equation}
m_W = a_tm + \frac1{2\xi} \hat p^2 + 2\sin^2\left(\frac{\theta^a-\theta^b}2\right).
\end{equation}
To proceed we write 
\begin{eqnarray}
V(\Omega) &=&  2\,{\rm Re} \, \sum_{a\neq b} \, \sum_{r=1}^\infty V_r \, e^{ir\left(\theta^a-\theta^b\right)} + {\rm constant} \nonumber \\
&=& 2\,\sum_{r=1}^{\infty}\, V_r\, \left|{\rm tr}\, \Omega^r \right|^2 + {\rm constant},
\end{eqnarray}
with
\begin{eqnarray}
V_r &=& {\rm Re} \, \int \frac{d\omega}{2\pi}\,e^{ir\omega}\, \int \left(\frac{dp}{2\pi}\right)^3\, \left\{\log \left[ \frac1{\xi^2}\, \hat p^2 + 4\sin^2\left(\frac{\omega}{2}\right)\right]\right. \nonumber \\
&&\left. - 2N_f \log\left[   \frac1{\xi^2}\, \hat{\hat{p}}^2 + \sin^2\left(\omega\right) + m^2_W(p,\omega)\right]\right\}.\label{1loop1}
\end{eqnarray}
In Appendix~\ref{appA} we show that $V_r$ is given by
\begin{equation}
V_r = \frac1{r} \int_{-\pi}^\pi \left(\frac{dp}{2\pi}\right)^3\,\left[ 2N_f \,e^{-r \, E_F(p)} - e^{-r\, E_G(p)}\right], \quad r\ge 1,
\label{1loop2}
\end{equation}
with $E_F(p)$ and $E_G(p)$ the dispersion relations of the fermions and gluons on the lattice
\begin{eqnarray}
E_G(p) &=& 2\sinh^{-1}\left( \frac{\hat p}{2\xi}\right),\label{EG} \\
E_F(p) &=& 2\sinh^{-1}\left( \frac{1}{2\xi} \sqrt{\frac{\hat{\hat p}^2  + \left(\xi\, a_tm + \frac1{2} \hat p^2\right)^2}{1 + a_tm + \frac{1}{2\xi} \, \hat p^2}}\right).
\label{EF}
\end{eqnarray}
We note that, in physical units, the potential $V(\Omega)$ (or its Fourier components $V_r$) are multiplied by $1/(a_t a^3_s) = 1/(\xi^3a_t^4)$. To avoid confusion we shall denote the (dimensionful) physical potential and its Fourier components by a superscript `phys'.

\subsection{The meaning of $r$, the embedding of spacetime into color space, and why should there be a cascade of transitions with increasing fermion mass.}
\label{meaning}

Eguchi-Kawai spacetime reduction can be justified in perturbation theory. To see this one performs weak coupling calculations in the field theory and in the reduced model, and shows that the planar diagrams of the two theories are the same \cite{PT}. Since the fields in the volume-reduced theory have no spacetime coordinates, and consequently no momentum, one may wonder how can such an equivalence be valid. Put differently, what is it that flows  along the lines in the Feynman diagrams of the reduced theory?

The answer is tied with the fact that, in the reduced model, perturbation theory  is performed around a particular ground state which is $Z_N$ invariant. This means that the zero modes --- the classical values of the gauge fields (or the logarithm of the Polyakov loops) --- have eigenvalues that are uniformly distributed along the unit circle. If we denote the eigenvalues of these matrices by $\theta^a_\mu$ (here $a$ is a color index and $\mu$ the index of the euclidean direction -- we consider the generic case where more than one direction is compactified), then the propagator of the fluctuating adjoint fields in the reduced model (the gluons and the adjoint fermions) would depend on $\theta^a_\mu$. This dependence can be shown to be a very special one. The combination  $(\theta^a_\mu - \theta^b_\mu)$ behaves like the $\mu^{\rm th}$ momentum component 
of the $(ab)^{\rm th}$ color component of the adjoint field. This is anticipated since in a gauge theory a momentum variable is replaced by a covariant momentum, and, in the presence of a classical background of the form $A^{ab}_\mu \propto \delta_{ab}\,\theta^a_\mu$, the covariant momentum of an adjoint particle is indeed $(\theta^a_\mu - \theta^b_\mu)$. If the ground state is $Z_N$ invariant, then, in the large-$N$ limit, the latter difference of angles would be uniformly distributed between $-\pi$ and $+\pi$, exactly what we expect from a lattice momentum variable.

The fact that $(\theta^a_\mu - \theta^b_\mu)$ behaves like a momentum variable can also be seen from \Eq{1loop1}. There, this difference is replaced by the variable $\omega\in [-\pi,\pi)$,  and we see that the inverse gluonic and fermionic propagators (the arguments of the logarithms)  are precisely the propagators of fields defined on $R^4$ and carrying a euclidean four momentum equal to $(\vec p,\omega)$. Since the winding number $r$ is the conjugate to $\omega$ (again see \Eq{1loop1}), then $r$ behaves like a euclidean distance along an {\em uncompactified} dimension which replaces the single site that was the $S^1$.

The fact the winding of the Polyakov loops takes the role of a euclidean distance is most clearly seen from \Eq{1loop2}. There $r$ multiplies the energies of the gluons and fermions. In fact, in that equation, the different contributions to  $V_r$ look like correlation functions of gluons and fermions that are separated distance $r$ in euclidean space.

Once we understand that the coefficients $V_r$  are determined by the difference between fermion and gluon correlators along a euclidean distance $r$, we can anticipate the cascade of transitions mentioned above. To see this note that when the fermions have a nonzero mass and are heavier than the gluons, they propagate a shorter distance -- their contribution at large enough $r$ is negligible compared to that of the gluons and only the latter will determine the behavior of highly wound loops. Thus, as we increase the fermion mass, lower and lower winding of loops become determined solely by the gluons. Since  we know gluons generically cause the winding Polyakov loops to condense (indeed observe the minus sign of the gluonic contribution in \Eq{1loop2}), we anticipate a cascade of transitions, where lower and lower winding of loops condense with increasing fermion mass. 

\section{The minimum of the one-loop potential}
\label{minimum}
In this section we minimize the one-loop potential. In \Sec{m0_lat} and
 \Sec{m0_cont} we prove analytically that at $m=0$ and for $0\le \xi<2$ the ground state is $Z_N$ invariant. Then, in \Sec{m_nonzero} we numerically calculate the values of $V_r$ for nonzero mass, for $N_f=1/2,1,2$, and for values of $\xi$ in the range $[0,4]$.

\subsection{The massless case on the lattice}
\label{m0_lat}
In this section we ask what is the sign of $V_r$ when $m=0$. We do so by comparing $E_F$ and $E_G$. From \Eq{1loop2} we see that $V_r$ will be positive for all $r\ge 1$ if $N_f\ge 1/2$ and $E_G(p)\ge E_F(p)$. From Eqs.~(\ref{EG})--(\ref{EF}) we see that the latter inequality will be fulfilled when 
\begin{equation}
\hat p^2 \ge  \frac{\hat{\hat{p}}^2 + \frac14 \, \hat p^4}{1 + \frac1{2\xi} \hat p^2}.\label{ineq1}
\end{equation}
Using the identity $\hat{\hat{p}}^2= \hat p^2 - 4 \sum_{i=1}^3 \sin^4\left(p_i/2\right)$, this inequality  becomes
\begin{equation}
\frac14\,\hat p^4 \left(\frac{2}{\xi} - 1\right) \ge 
 -4 \sum_{i=1}^3 \sin^4\left(p_i/2\right), \label{ineq2}
\end{equation}
which holds if $0\le \xi \le 2$. Therefore, in this regime one finds that for $N_f \ge \frac12$ 
\begin{equation}
V_r \ge 0.
\end{equation}
In particular, for one or more flavors we get $V_r >0$ which means that $V(\Omega)$ is minimized when ${\rm tr} \, \left[\Omega^r\right]=0$ for $|r|\ge 1$, and the ground state is $Z_N$ symmetric.

\subsection{The massless case in the spatial continuum limit of $a_s\to 0$ and $a_t$ fixed.}
\label{m0_cont}
Let us check what happens in the spatial continuum limit, where we keep $a_t$ fixed, but send $a_s$ to zero. In Appendix~\ref{appB} we show how to take the $\xi\to 0$ limit of \Eq{1loop2}. We find that the values of $V_r$ for $r=1,2,3$ have quadratic, linear, and logarithmic UV sensitivities, respectively. The origin of such divergences was explained in \Ref{one_loop} and calls for the addition of counter terms to the one-loop potential that are of the form $\left|{\tr \Omega}^r\right|^2\, ; \,r=1,2,3$. These counter terms can always be chosen such that the renormalized $V_{r=1,2,3}$ obey
\begin{equation}
V_{1,2,3}>0. \label{V123}
\end{equation}
Indeed, from the general form of \Eq{1loop2} with $E_F(\vec p)=E_G(\vec p)\equiv E(\vec p)$, we see that for $N_f\ge 1$ the choice in \Eq{V123} is a most natural renormalization condition. The rest of the $V_r$'s are finite and in the massless case they have the following form.
\begin{equation}
V^{\rm phys.}_{r\ge 4} = \frac{2N_f-1}{\pi^2a^4_t}\, \int_{-\infty}^{\infty} \left(\frac{dp}{2\pi}\right)^3\, e^{-r\,E(\vec p)} \quad ;\quad E(p) = 2\sinh^{-1}|\vec p|.\label{1loopxi0}
\end{equation}
(Note that here the integration variable $p$ is dimensionless). In Appendix~\ref{appB} we show that \Eq{1loopxi0} is what one obtains if one chooses to use a continuum regulator for the spatial UV (for example, a hard cutoff or dimensional regularization). Interestingly, while the integral in \Eq{1loopxi0} is the correct $\xi\to 0$ limit of \Eq{1loop2} only for $r\ge 4$ (otherwise it is UV divergent), then we note that inspecting the UV behavior of the right hand side in  \Eq{1loopxi0} shows that it is finite for all $r\ge 2$, and UV linear divergent only for $r=1$. This linear divergence is one of the UV sensitivities of the one-loop potential that \Ref{one_loop} exposed by analyzing a continuum regulator.\footnote{Another divergence was argued to be present in the $r=2$ term, but here we see that when we regulate the compact dimension with Wilson fermions that have a `Wilson parameter' equal to unity, the coefficient of the linear divergence is exactly zero.\label{Vr2}} This means that if we regulate the spatial UV of the single site theory on a lattice, and then take the $\xi\to 0$ limit, then the UV sensitivity of $V^{\rm phys.}(\Omega)$ is more severe than what we find in a continuum regularization of the spatial UV. For further details on how to obtain \Eq{1loopxi0} from continuum regulators see Appendix~\ref{appC}.

\bigskip

 The simple form of the dispersion relation in \Eq{1loopxi0} allows us to analytically calculate the coefficients $V^{\rm phys.}_r$. We do so by inverting the dispersions, $|\vec p| = 2\sinh (E/2)$, and writing \Eq{1loopxi0} as 
\begin{equation}
V^{\rm phys.}_r =  \frac{(2N_f-1)}{\pi^2ra_t^4} \int_0^\infty dE \,\sinh (E/2) \, \sinh (E) \, e^{-rE}.\label{Vrcont}
\end{equation}
For $r\ge 2$, \Eq{Vrcont} leads to 
\begin{equation}
V^{\rm phys.}_r =  \frac{2N_f-1}{\pi^2a_t^4} \, \frac1{\left(r^2-\frac94\right)\left(r^2-\frac14\right)},\label{Vrcont1}
\end{equation}
which is positive for $N_f\ge 1$. Combining \Eq{V123} and \Eq{Vrcont1} we see that the one-loop potential for $N_f\ge 1$ and $m=0$ is minimal when the ground state respects the $Z_N$ symmetry. 

As a check of our calculation  we note in passing that for large $r$ the right hand side of \Eq{Vrcont1} is precisely the one that appears in the calculation of Kovtun-\"Unsal-Yaffe in \Ref{KUY} (if we identify the distance $a_t$ with the extent of the compact direction in that calculation).
This is expected from \Eq{1loop2}. There, we see that when $r$ is large, the spatial momenta that would contribute to the integral have a corresponding energy that obeys $E\lesssim 1/r \ll 1$. For such momenta we can replace $E$ by $2\sinh (E/2)=|\vec p|$, and we arrive at the continuum equations that appear in \Ref{KUY}. Thus, looking at large winding $r\gg 1$ leads  naturally to the continuum limit in the compact direction. We extend this observation to the massive case in Appendix~\ref{appD}, where we show that in the large-$r$ limit $V^{\rm phys.}_r$ agrees with the result of \Ref{MH}.

In Appendix~\ref{appC} we perform an additional check of our calculation and verify that  \Eq{Vrcont1} is obtained with a different mathematical procedure (i.e.~first integrating over $p$ and then over $\omega$) and within a different regulator for the spatial $R^3$ directions. Specifically, we compare with a hard cutoff and with dimensional regularization.

Finally, we note that the continuum limit of our results for massless quarks is also consistent with the single-site  case of \Ref{PU} (There the non-compact directions were regularized in the continuum, and the compact direction was allowed to consist of $\Gamma\ge 1$ sites).

\subsection{Minimizing the lattice one-loop potential for general values of $m$, $\xi$, and $N_f$.}
\label{m_nonzero}

\begin{figure}[hbt]
\centerline{
\includegraphics[width=18cm]{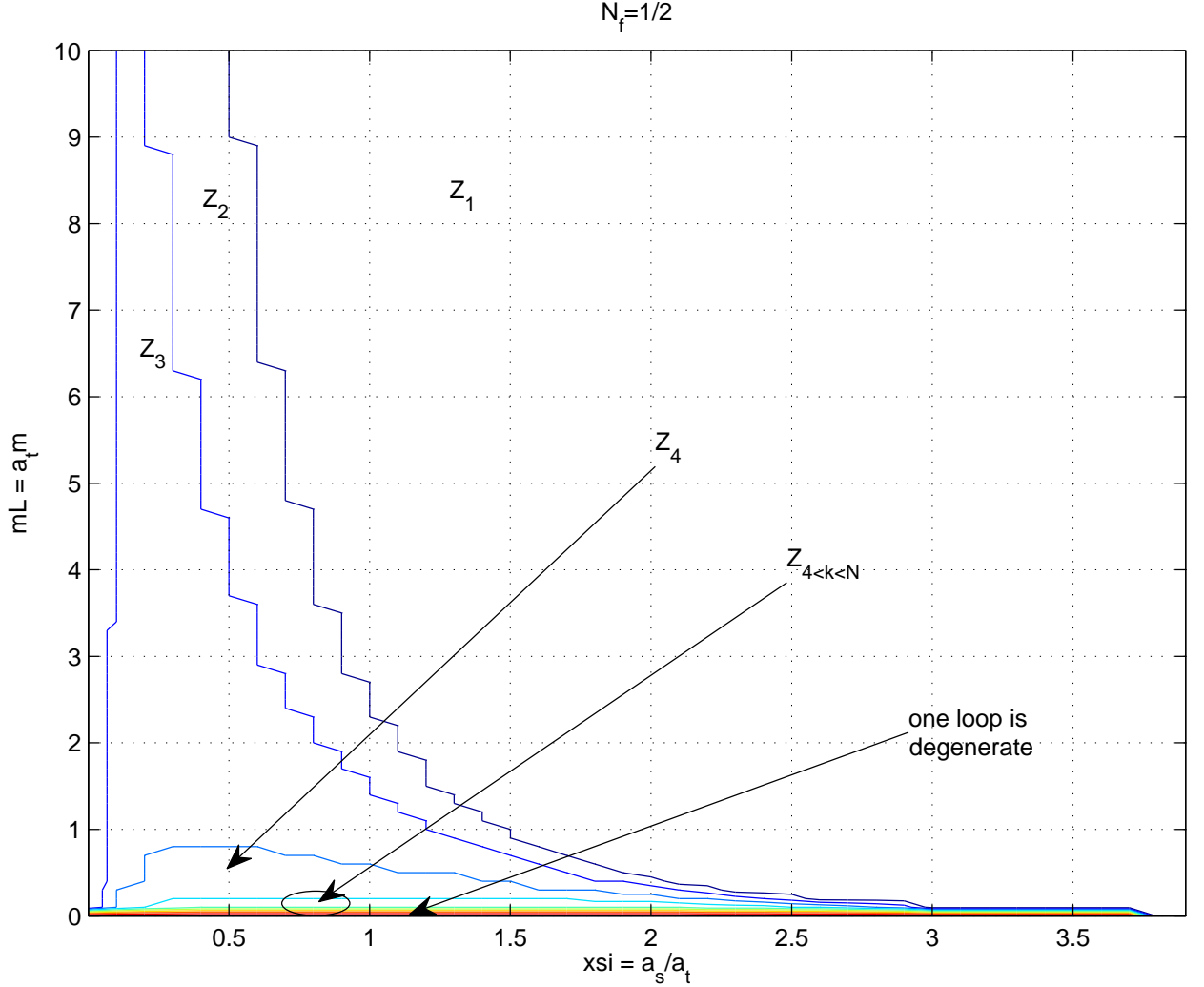}
}
\caption{Large-$N$ phase diagram
for theories with a single Majorana Dirac fermion.
}
\label{Nf05}
\end{figure}

\begin{figure}[hbt]
\centerline{
\includegraphics[width=18cm]{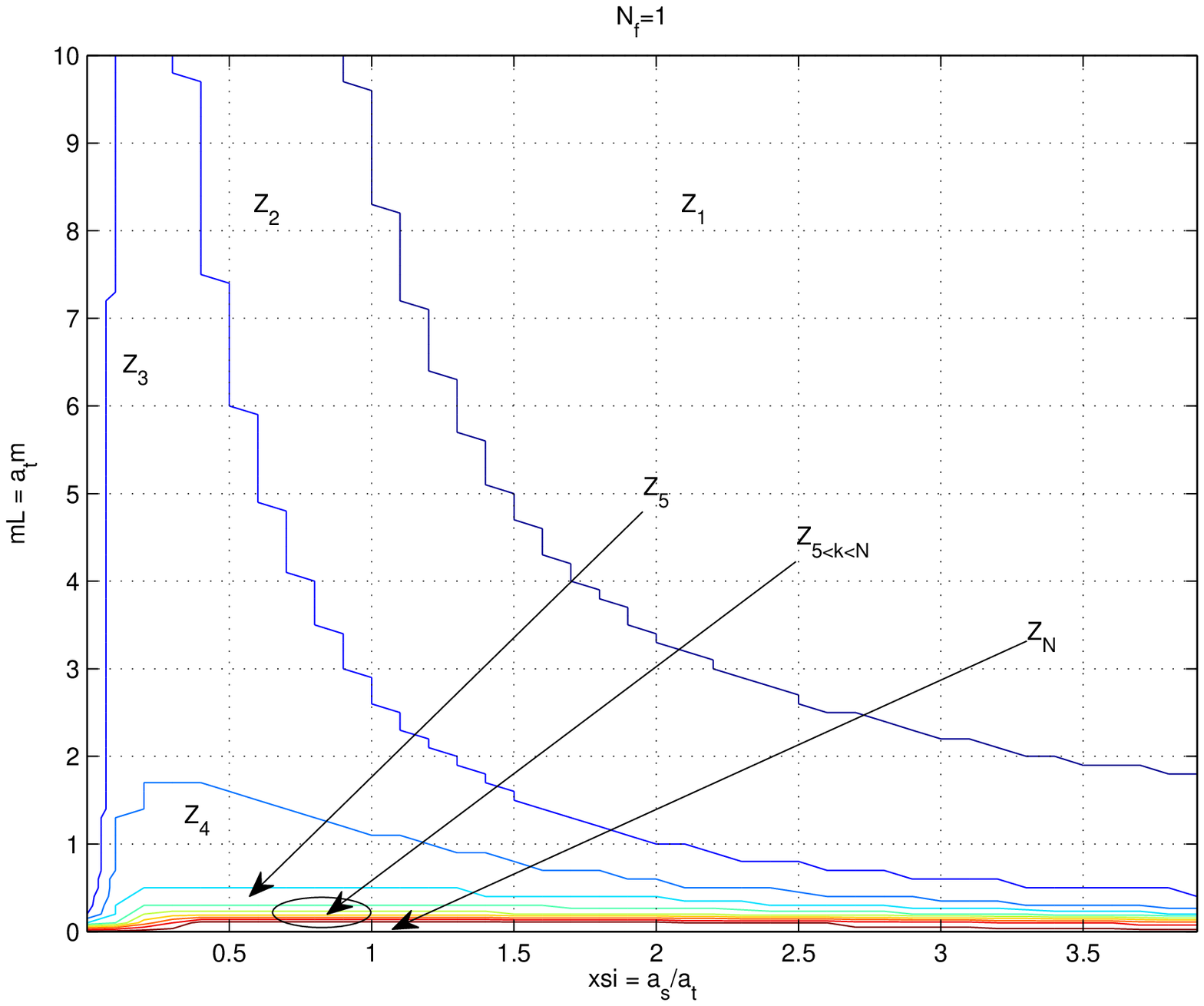}
}
\caption{Large-$N$ phase diagram
for theories with two Majorana fermions.
}
\label{Nf1}
\end{figure}

\begin{figure}[hbt]
\centerline{
\includegraphics[width=18cm]{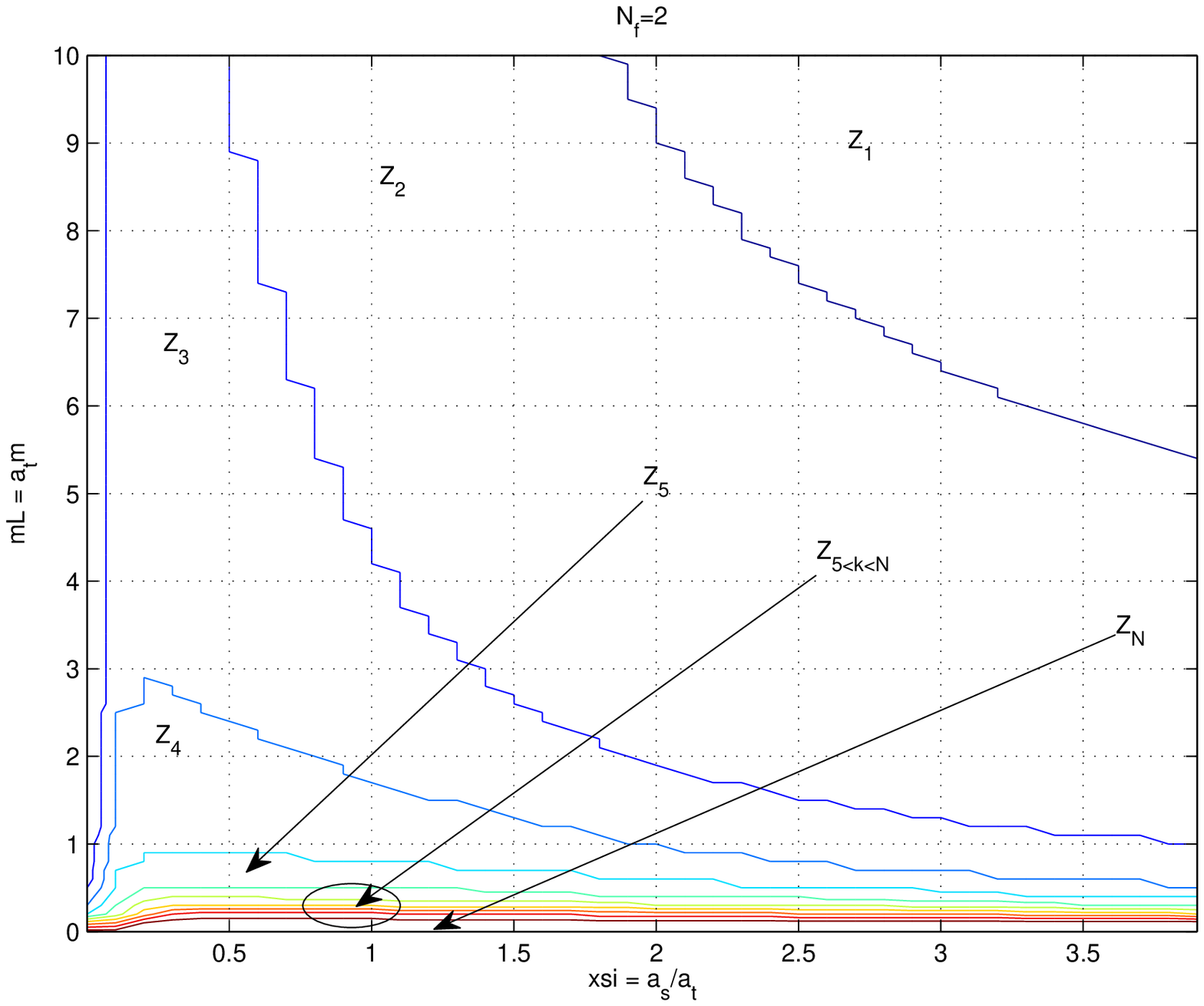}
}
\caption{Large-$N$ phase diagram
for theories with four Majorana fermions.
}
\label{Nf2}
\end{figure}

We calculated $V_r$ for general values of $a_tm$, $\xi$, and $N_f$ by numerically performing the spatial integrals in \Eq{1loop2}.\footnote{The numerical integrals were performed with a trapezoid method and a resolution of the interval $[-\pi,\pi)$ in each direction equal to $2\pi/L$ with $L=30,50,90,130$, and $150$. The results we present were stable under a change in $L$.}  For each choice of parameters we denote by $K$ the smallest value of $r$ for which $V_r$ is negative. For this parameter choice $Z_N$ breaks to $Z_K$. Since we numerically calculated $V_r$ for $r=1,2,3,\dots,20$, then, from these results, we can only conclude about symmetry breaking to subgroups that are at most $Z_{20}$. Nevertheless, in Appendix~\ref{appD} we investigate the large-$r$ limit and show that in that limit
the symmetry realization is identical  to the one of the continuum calculation done in \Ref{MH}. In particular, we see that the when $a_tm\ll 1$, there is an instability for $Z_N$ to break to $Z_K$ with $K\gg 1$ if
\begin{equation}
a_tm > c_{N_f}/K \quad ; \quad c_{1,2}\simeq 2.03,3.155. \label{m_c_cont}
\end{equation}
For generic values of $a_tm$, where the breaking of $Z_N$ is to $Z_K$ with $K\sim O(1)$, the values of the lattice $V_r$'s (and the resulting phase diagram) can be obtained by numerical integrations of \Eq{1loop2}, and we present our results in Figs.~(\ref{Nf05}--\ref{Nf2}). 

As the figures show, for all values of $\xi$ we find that the ground state at $m=0$ has the highest possible symmetry. Since we restricted our numerical study to instabilities of $Z_N$ breaking to $Z_r$ with $r \le 20$, we cannot conclude whether $Z_N$ breaks to $Z_r$ with $r>20$, but from \Eq{m_c_cont} we see that at large enough values of $r$ and at $a_tm=0$, then the $Z_N$ is intact for all $\xi$.

When $a_t m$ increases from zero, the $Z_N$ symmetry breaks to a subgroup $Z_r$ with $r$ a decreasing function of $a_tm$. We emphasize that while the figures seem to indicate that there is a finite range of $a_tm$ for which the $Z_N$ in intact (see white patch just above the $x$-axis in Figs.~(\ref{Nf1}--\ref{Nf2})), then this is most likely  an artifact of our restriction to investigate numerically symmetry breakdowns of the form $Z_N\to Z_r$ with $r\le 20$. In particular, as we mention above (see \Eq{m_c_cont}), in Appendix~\ref{appD} we see that even for infinitesimal values of $a_tm$, there is an instability for the  $Z_N$ symmetry to break to a subgroup $Z_r$ with $r \sim (a_tm)^{-1}$.

\section{Summary}
\label{summary}

The calculation in this paper is concerned with large-$N$ lattice QCD with adjoint fermions on $R^3\times S^1$. We calculate the one-loop potential $V$ of this theory as a function of the Polyakov loop $\Omega$ that wraps the compact direction. We use  Wilson fermions and the compact direction consists of a single site. Our calculation extends \Ref{one_loop} by that it directly determines, for different values of the quark mass, the lattice anisotropy parameter $\xi$, and the number of flavors, $N_f$, the instabilities in $V(\Omega)$ toward the breakdown of the $Z_N$ center symmetry to a $Z_K$ subgroup with $K<N$.

Our results are summarized below.
\begin{itemize}
\item For $m=0$ we prove analytically that the minimum of the one-loop potential respects the $Z_N$ symmetry. This is true for $N_f\ge 1$ and in the regime $0\le \xi<2$, which includes the case where one takes the continuum limit in the uncompactified $R^3$ directions.
\item For $m=0$ and  $N_f=1/2$ (a single Majorana fermion) the one loop potential is zero in the continuum  limit of the uncompactified directions and so one cannot use it to determine the ground state. For any finite $a_s$ the one loop potential is nonzero and prefers a $Z_N$ symmetric ground state.\footnote{For a discussion on the nonperturbative effects of instantons see \Ref{KUY}.}
\item For $m>0$ we see a cascade of transitions where the $Z_N$ breaks to a smaller $Z_K$ subgroup with $K<N$, and $K$ being a decreasing function of $a_tm$. At $a_tm =\infty$ the center $Z_N$ completely breaks down.
\item When we investigate the possibility to break $Z_N$ to $Z_K$ with $K\gg 1$ with the {\em lattice} one-loop potential, we find the {\em continuum} results of \Ref{MH}: when $m$ is increased to values of $O(1/K)$, then $Z_N$ breaks to $Z_K$. Thus the Polyakov loops with large winding numbers behave like they do in the continuum.
\end{itemize}
The fact that we see symmetry breakdown of $Z_N$ for a wide range of masses means that one of the conclusions of \Ref{one_loop} does not hold. There, we concluded that the $Z_N$ symmetry is intact even for moderately heavy masses, but relied on comparing the one-loop potential only for ground states that either leave $Z_N$ intact or break it down to $\O$ or $Z_2$. Here we see that for intermediate masses the symmetry breaks to $Z_r$ subgroups.

We wish to emphasize that, while our results are very similar to those obtained in the continuum works Refs.~\cite{KUY,MO,MH}, then the UV sensitivities of the one-loop potential discussed in \Ref{one_loop} make it impossible to determine a priory whether this similarity will take place or not (this is because our lattice theory has only one site in the compactified direction).

The methods we use make clear that the winding number of the Polyakov loop around the compact direction is identified with a euclidean distance along the direction one reduces (at least in the single-site case that we studied in this paper). As we explain in \Sec{meaning}, this leads one to anticipate the cascade of transition seen in our calculation and in Refs.~\cite{MO,MH}: when $m>0$ the fermions are heavier than the gluons and the winding-number/distance identification tells us that their effect will propagate to smaller number of windings than that of the gluons. Therefore, at nonzero $m$, the behavior of large enough winding numbers will be determined by gluons, and these will cause a condensation of the multiply wound Polyakov loops, leading to a partial breakdown of the $Z_N$ center.

Finally, we wish to remark on the complications of inferring from the results obtained in this paper on nonperturbative lattice simulations. First, we work at weak coupling, and it is not clear how the stronger fluctuations induced by  stronger lattice couplings (where one would actually perform lattice simulations) change the one-loop picture. Second, our spacetime geometry is that of $R^3\times S^1$, while standard lattice simulations are often done on a four-torus. Indeed, the one-loop potential for a system that has more than one reduced direction depends on more than one Polyakov loop. Furthermore, if one defines the theory on a lattice with a  single site in all directions, then the one-loop calculation will not be valid; there are zero modes around specific vacua (`singular tolerons' -- see \Ref{0modes}) that cause IR divergences and require a  nonperturbative treatment. 

The last difficulty that arises when one tries to infer from our results on what would happen in simulations is related to $1/N$ corrections. Here we note that when we minimized the one-loop potential we assumed that all the windings of the Polyakov loops are independent; an assumption that holds only when $N=\infty$. At finite values of $N$  the minimization is more complicated and, for the continuum one-loop potential, was performed numerically in Refs.~\cite{MO,MH}. The result of these references is interesting: if we denote the size of the $S^1$ by $L$, then one finds that there is a critical value of $(m\times L)\equiv \mu_c$, below which the cascade of transitions ends and the $Z_N$ symmetry is completely intact. In particular,  just above $\mu_c$ the $Z_N$ breaks to its largest subgroup, $Z_{N/2}$. Thinking about this breakdown as $Z_N\to Z_K$ with $K\gg 1$, we can anticipate that $\mu_c(N) \sim 1/N$ (see last bullet above). Thus, while for given values of $N$, $L$, and $m$ that obey $\mu < \mu_c$, the $Z_N$ might be intact, then, according to the $R^3\times S^1$ one-loop analysis, an increase in $N$ (done at fixed $m$ and $L$) will result in a breakdown of $Z_N$. Since lattice simulations are always done at finite value of $N$, then the breakdown we discuss above and the cascade of transitions we discuss in this paper, might occur only if one systematically checks how the simulation results change when $N$ is increased. Such systematic checks were done in \Ref{BS} to the extent possible with the metropolis  simulation algorithm used in that paper (in practice this algorithm allowed the study of $8\le N \le 15$). It will be very useful to develop faster algorithms and simulate larger value of $N$, so to extend these systematic checks further.

The caveats mentioned above might explain why the nonperturbative lattice studies presented in \Ref{CD,BS,HN} observed a $Z_N$ symmetric ground state even at nonzero, but small, mass.\footnote{While \Ref{CD} finds an elaborate phase structure for $SU(3)$, it reports a center invariant phase at sufficiently weak couplings.} \footnote{\Ref{CD} and \Ref{HN} used staggered and overlap fermions, respectively, and so the results we obtained in this paper do not directly related to that work (the one-loop potentials on lattices with a small number of points sometimes have UV sensitivities that requires the addition of certain relevant operators when comparing  results obtained with different lattice actions.)}

\section*{Acknowledgments}

I thank A.~Armoni for pointing my attention to the way the results of \Ref{MH} behave on an $R^3\times S^1$ geometry. I also thank M.~Hanada, T.~Hollowood, J.~Myers, S.~R.~Sharpe, and M.~\"Unsal.
This work was supported by the U.S. Department of Energy under Grant No. DE-FG02-96ER40956.

\section*{Note added}
For $m=0$, the recent preprint \cite{PU} by Poppitz and \"Unsal is studying a similar construction to the one we study in this paper. Specifically, \Ref{PU} defined the theory to lie in $R^D\times S^1$, and regularized the UV of the $R^D$ in the continuum and the UV of the $S^1$ on a lattice of $\Gamma$ sites. For the case $D=3$ and $\Gamma=1$, this construction becomes the one we study in \Sec{m0_cont} --- i.e.~the limit in which we send the lattice spacing in the non-compact directions to zero. Indeed, in that limit, our formulas for $V_r$ (see e.g.~\Eq{Vrcont1} and those in appendix~\ref{appC}) become consistent with those of \Ref{PU}.

\appendix

\section{The calculation of $V_r$.}
\label{appA}

In this appendix we show how to arrive at \Eq{1loop2}. We begin by simplifying the argument of the fermionic logarithm of \Eq{1loop1} and write
\begin{eqnarray}
\frac1{\xi^2}\hat{\hat{p}}^2 + \sin^2 \omega + m^2_W(p,\omega) &=& 4\sin^2 \omega/2 \left(1 + a_tm + \frac1{2\xi} \hat p^2\right) + \frac1{\xi^2}\,\hat{\hat{p}}^2 + \left(a_t m + \frac1{2\xi} \hat p^2\right)^2,\\
&& = \left(1 + a_tm + \frac1{2\xi} \hat p^2\right) \left\{ 4\sin^2 \omega/2  +  \frac{\frac1{\xi^2}\,\hat{\hat{p}}^2 + \left(a_t m + \frac1{2\xi} \hat p^2\right)^2}{ \left(1 + a_tm + \frac1{2\xi} \hat p^2\right)} \right\}.\nonumber\\
\end{eqnarray}
We factored out  $\left(1+a_tm+\frac12 \hat{p}^2 \right)$ because it is independent of $\omega$ and does not contribute to $V_r$ when $r\ge 1$. We can now write
\begin{eqnarray}
V_r&=&{\rm Re}\, \int \left(\frac{dp}{2\pi}\right)^3\, \int \frac{d\omega}{2\pi}\,e^{ir\omega}\,\left\{\log \left[ 4\sin^2\left(\omega/2\right) + 4\sinh^2 \left(E_G(p)/2\right)\right]\right.\nonumber \\
&& \left.-2N_f \,\log\,\left[ 4\sin^2\left(\omega/2\right) + 4\sinh^2 \left(E_F(p)/2\right)\right] \right\}, \qquad {\rm for}\qquad r\ge 1.\label{1loop3}
\end{eqnarray}
with $E_{F,G}(p)$ the lattice dispersion relations of the fermions and gluons that we give in \Eq{EG} and \Eq{EF}. Defining $I_r[E]$ as 
\begin{equation}
I_r[E] = {\rm Re}\int \frac{d\omega}{2\pi}\, e^{ir\omega} \, \log \left[4\sin^2 \omega/2 + 4\sinh^2E/2\right].
\end{equation}
we have 
\begin{equation}
V_r=\int \left(\frac{dp}{2\pi}\right)^3 \left\{I_r[E_G(p)] - 2N_f I_r[E_F(p)]\right\}.\label{Ir1}
\end{equation}
To proceed, we need to calculate $I_r[E]$. We do so by first integrating by parts, then performing the change of variables $z=e^{i\omega}$, and finally using the Cauchy integral theorem:
\begin{eqnarray}
I_r(E) &=&  {\rm Re}\int \frac{d\left(e^{ir\omega}\right)}{2\pi ir}\, \log \left[4\sinh^2 E/2+ 4\sin^2 \omega/2 \right]\nonumber \\ 
&=& {\rm Re}\left(\frac{e^{i\omega r}}{2\pi ir} \log \left[4\sinh^2 E/2+ 4\sin^2 \omega/2 \right]^{2 \pi}_0 - \int_0^{2\pi} \frac{d\omega}{2\pi i r}\, e^{ir\omega}\, \frac{2\sin\omega}{4\sinh^2E/2+4\sin^2\omega/2}\right)\nonumber\\
&=& {\rm Re}\left(- \frac1{r}\,\oint\frac{dz}{2\pi i}\, \frac{z^{r+1}-z^{r-1}}{(z-e^{-E})(z-e^E)}\right) = -\frac{e^{-rE}}{r}.\label{Ir2}
\end{eqnarray}

Combining \Eq{Ir2} and \Eq{Ir1} gives \Eq{1loop2}.

\section{The $\xi\to 0$ limit of \Eq{1loop2}}
\label{appB}

In this technical appendix we show how to take the $\xi\to 0$ limit of \Eq{1loop2}. To do so we divide the integration over the Brillouin Zone into two regimes. 
\begin{enumerate}[{\rm Regime} I.]
\item The neighborhood of the Brillouin Zone origin. Here $\hat p \approx \hat{\hat{p}} \approx |\vec p| \equiv \xi\, q$ with $q\sim O(1)$. It is straight-forward to show that the contribution of regime~(I) to the one-loop potential is given by \Eq{1loopxi0}:
\begin{equation}
V^{\rm phys.}_{r\ge 4} = \frac{2N_f-1}{\pi^2a^4_t}\, \int_{-\infty}^{\infty} \left(\frac{dp}{2\pi}\right)^3\, e^{-r\,E(\vec p)} \quad ;\quad E(p) = 2\sinh^{-1}|\vec p|.\label{1loopxi01}
\end{equation}
\item The rest of the Brillouin Zone, where  $\hat p\sim O(1)$. In that regime we have
\begin{eqnarray}
E_G &\stackrel{\xi\to 0}{\longrightarrow}&2 \log \left(\hat p/\xi\right),\label{EG1}\\
E_F &\stackrel{\xi\to 0}{\longrightarrow}& \log \left(\left(\frac{2\hat{\hat p}^2}{\hat p^2}+\frac12 \hat p^2\right)/\xi\right).\label{EF1}
\end{eqnarray}
Substituting Eqs.~(\ref{EG1}--\ref{EF1}) into \Eq{1loop2} (and dividing $V_r$ by $\xi^3a^4_t$ to get $V^{\rm phys.}_r$) we find
\begin{equation}
V^{\rm phys.}_r \stackrel{{\rm regime\,(II)}}{\sim} \frac1{a^4_t}\, \left[  2N_f\, {\cal O}\left(\xi^{r-3}\right) \, + \, {\cal O}\left(\xi^{2r-3}\right)  \right].\label{regimeII}
\end{equation}
Where here the first term is the contribution of the fermions, and the second of the gluons.
\end{enumerate}

Combining Eqs.~(\ref{1loopxi01}) and~(\ref{regimeII}) we find that 
\begin{itemize}
\item  The gluons render the $V_{r=1}$ linearly divergent.
\item The fermions render $V_{r=1,2,3}$ quadratically, linearly, and logarithmically, divergent, respectively.
\item Only $V_{r \ge 4}$ are finite in the $\xi\to 0$ and are given by \Eq{1loopxi0}.
\end{itemize}

\section{Comparison of $V^{\rm phys.}_r$ in the $r\to \infty$ limit with the result of \Ref{MH}.}
\label{appD}

In this appendix we show how, for large values of $r$, the coefficients $V_r$ in the lattice one-loop potential, become those obtained previously in the continuum calculation of \Ref{MH}. This serves as a check of our results and also allows us to conclude that for small, but finite, values of $m$, there is an instability for the $Z_N$ to break to $Z_K$ with $K\sim 1/m \gg 1$. 

Our starting point is to take the large-$r$ limit of \Eq{1loop2}. For $r\gg 1$, the energies $E_{F,G}(p)$ that contribute to $V_r$ are those that obey
\begin{equation}
E_{F,G}(p) \lesssim 1/r \ll 1.
\end{equation}
For such small energies we can write
\begin{eqnarray}
E_G(p) &=& \left|\vec p\right|/\xi,\label{EG1} \\
E_F(p) &=& \left(\left|\vec p \right|^2 + (\xi a_tm)^2\right)^{1/2}/\xi\label{EF1}.
\end{eqnarray}
Substituting Eqs.~(\ref{EG1}--\ref{EF1}) into \Eq{1loop2} and using straight-forward manipulations gives
\begin{equation}
V^{\rm phys.}_{r\gg 1} =\frac1{a^4_t\pi^2r^4} \left(N_f \, \mu^2_r \, K_2(\mu_r) -1\right)\quad ; \quad \mu_r = r\, a_tm
.\label{1loop21}
\end{equation}
Identifying $a_t$ with the length $L$ of the compactified direction in \Ref{KUY} and \Ref{MH}, we see that the large-$r$ result in \Eq{1loop21} becomes identical with the results of these works.

The meaning of \Eq{1loop2} is that when the expression in the brackets becomes negative, there is an instability of the $Z_N$ symmetry to break to $Z_r$. This will happen when $\mu_r$ obeys
\begin{eqnarray}
\mu_r &\gtrsim&  2.03 \quad N_f=1\\
\mu_r &\gtrsim& 3.155 \quad N_f=2. 
\end{eqnarray}
This gives \Eq{m_c_cont}.

\section{Comparison of the result obtained  in the continuum limit of the uncompactified directions (\Eq{Vrcont1}) with other regulators of the spatial UV.}
\label{appC}

As a check we compare the result obtained in \Eq{Vrcont1} with the one obtained employing a continuum regulator for the spatial $\int d^3p$ integrals. The regulator we choose is a hard cutoff in momentum space, but we also remark below on what one would obtain with  minimally subtracted dimensional regularization.

When the spatial space is a continuum, the one-loop potential of the massless theory is given by (here the factor of $1/a_t$) is required to make the potential dimension four)
\begin{equation}
V(\Omega) = \frac{1-2N_f}{a_t} \sum_{a\neq b}\, I\left[\frac4{a^2_t}\sin^2\left(\frac{\theta^a-\theta^b}{2}\right)\right],\label{Vcont}
\end{equation}
where $I\left[M^2\right]$ is 
\begin{equation}
I\left[M^2\right] = \int \left(\frac{dk}{2\pi}\right)^3\, \log \left[ k^2 + M^2\right].
\end{equation}
This integral has both cubic and linear divergences. The cubic ones can be removed by subtracting $I(0)$, and with a hard cutoff we obtain
\begin{eqnarray}
I(M^2) - I(0) &=& \int \left(\frac{dk}{2\pi}\right)^3\, \log \left[1 + \frac{M^2}{k^2}\right] = \frac1{2\pi^2}\int k^2dk\log  \left[1 + \frac{M^2}{k^2}\right]\nonumber\\
&=&\frac1{6\pi^2}\int d(k^3)\,\log  \left[1 + \frac{M^2}{k^2}\right]=\frac1{6\pi^2}\, \left\{\left(k^3\log  \left[1 + \frac{M^2}{k^2}\right]\right)_0^{\Lambda} \right.\nonumber\\
&&\left. - \int dk\, \frac{k^3M^2(-2)\,k^{-3}}{1+\frac{M^2}{k^2}}\right\}=\frac{\Lambda M^2}{6\pi^2} + \frac{M^2}{3\pi^2}\int \frac{dk}{1+\frac{M^2}{k^2}}\nonumber\\
&=&\frac{\Lambda M^2}{6\pi^2}+ \frac{\left| M\right|^3}{3\pi^2}\int_0^{\Lambda/|M|} \frac{dx}{1+1/x^2} =  \frac{\Lambda M^2}{6\pi^2}+ \frac{\left| M\right|^3}{3\pi^2}\int_0^{\Lambda/|M|}dx \nonumber\\
&&- \frac{\left| M\right|^3}{3\pi^2}\int_0^{\Lambda/|M|} \frac{dx}{1+x^2}=  \frac{\Lambda M^2}{6\pi^2}+ \frac{\Lambda M^2}{3\pi^2} - \frac{\left|M\right|^3}{3\pi^2}\tan^{-1}\left(\Lambda/M\right)\nonumber\\
&\stackrel{\Lambda/M \to \infty}{\longrightarrow}&\frac{\Lambda M^2}{2\pi^2} - \frac{\left|M\right|^3}{6\pi}.\label{Vcutoff}
\end{eqnarray}
In the case of dimensional regularization (which implicitly sets the linear divergence to zero) we get 
\begin{equation}
I(M^2) \stackrel{\rm dim-reg.}{=} -\frac{\left|M\right|^3}{6\pi}.
\end{equation}

Let us now calculate the moments $V_r$ of $V(\Omega)$. To do so we start from \Eq{Vcutoff} which we write as (here we drop constants that do not depend on $\theta$)
\begin{eqnarray}
V(\Omega)&=&-\frac{(1-2N_f)}{2\pi a^4_t}\,\sum_{a\neq b}\, \left[\frac{2a_t\Lambda}{\pi} \cos\left(\theta^a-\theta^b\right) + \frac{8}3 \left|\sin\left(\frac{\theta^a-\theta^b}{2}\right)\right|^3 \right]\nonumber \\
&&=-\frac{(1-2N_f)}{2\pi a^4_t}\,\left[\frac{2a_t\Lambda}{\pi} \left|{\rm tr} \, \Omega\right|^2 + \sum_{a\neq b}\frac{8}3 \left|\sin\left(\frac{\theta^a-\theta^b}{2}\right)\right|^3 \right].\label{Vcutoff1}
\end{eqnarray}
If we write
\begin{equation}
\left|\sin\left(\omega/2\right)\right|^3 = 2\sum_{r=1}^{\infty}\, W_r\, \cos\left(\omega r\right) + W_0,
\end{equation}
then we have 
\begin{equation}
V_r = \frac{2N_f-1}{2\pi a^4_t}\, \left[ 
\begin{array}{lr}
\frac{a_t\Lambda}{\pi} + \frac{8}3 W_r, & \quad {\rm for}\quad r = 1.\\
\frac{8}3 W_r &  \quad {\rm for} \quad r\ge 2.
\end{array}
\right.
\end{equation}
The calculation of $W_r$ involves a standard Fourier transform:
\begin{eqnarray}
W_r &=& {\rm Re}\, \int_0^\pi \frac{d\omega}{\pi}\, e^{ir\omega}\, \sin^3\omega/2 = -\frac{1}{8\pi}\,{\rm Re}\, \int d\omega\, \,e^{ir\omega}\, \,\frac{e^{3i\omega/2} - 3e^{i\omega/2}+3e^{-i\omega/2}-e^{-3i\omega/2}}{i}\nonumber \\
&=&-\frac{1}{8\pi}\,{\rm Re}\,\left[\frac{-i(-1)^r-1}{i^2(r+3/2)}- 3\frac{i(-1)^r-1}{i^2(r+1/2)}+3\frac{-i(-1)^r-1}{i^2(r-1/2)}-\frac{i(-1)^r-1}{i^2(r-3/2)}\right]\nonumber \\
&=& -\frac{1}{8\pi}\left[\frac{-3}{r^2-9/4}+\frac{3}{r^2-1/4}\right]=+\frac{3}{4\pi}\,\frac{1}{(r^2-\frac94)(r^2-\frac14)}.\label{Wr}
\end{eqnarray}
which gives
\begin{equation}
V_r = \frac{(2N_f-1)}{\pi^2 a^4_t}\, \left[ 
\begin{array}{lr}
\frac{a_t\Lambda}2 - \frac{16}{15}, & \quad {\rm for}\quad r = 1,\\
\frac{1}{(r^2-\frac94)(r^2-\frac14)} &  \quad {\rm for} \quad r\ge 2.
\end{array}
\right. \label{Vr12}
\end{equation}
\Eq{Vr12} is precisely what we obtained in \Eq{Vcont} for $r\ge 2$.
 
The fact that $V_1$ is UV sensitive is explained in \Ref{one_loop} and means that the naive continuum limit one takes to arrive at \Eq{Vcont} requires a counter term. This counter term can be chosen such that the renormalized $V_1$ has arbitrary sign, and so such choice can affect the realization of the $Z_N$ symmetry. It also indicates that the $Z_N$ realization of the lattice theory can depend on the particular action one chooses to work with. For further discussion on this point see \Ref{one_loop}.

\end{document}